\documentstyle[11pt,newpasp,twoside,epsf]{article}
\def\edcomment#1{\iffalse\marginpar{\raggedright\sl#1\/}\else\relax\fi}
\reversemarginpar
\begin{document}

\title{Compressional Heating of
Accreting White Dwarfs in CV's}

\author{Dean M. Townsley}
\affil{Department of Physics, Broida Hall 
University of California, Santa Barbara, CA 93106}
 
\author{Lars Bildsten} 
\affil{Institute for Theoretical Physics and Department of Physics\\
Kohn Hall, University of California, Santa Barbara, CA 93106}

\begin{abstract}

 In recent years several Dwarf Novae (DN) systems have been observed
in quiescence, when the accretion rate is low and the WD photosphere
can be directly detected. The WDs are observed to cool after the DN
outburst from high effective temperatures to lower effective
temperatures ($T_{\rm eff}$) thought to be indicative of the thermal
state of the deep interior of the WD. Sion has argued that the most
likely energy source for this quiescent luminosity is the
gravitational compression of the WD interior, which rejuvenates an
otherwise cold WD into a much hotter state. We are undertaking a
theoretical study of the compressional heating of WD's, extending down
to the very low time averaged accretion rates, $\langle \dot M \rangle\sim
10^{-11}M_\odot \ {\rm yr^{-1}}$, applicable to the post-turnaround
CV's (the ``TOADS''). Nuclear burning is unstable at these $\langle \dot M \rangle$'s,
so we have incorporated the recurrent heating and cooling of the WD
throughout the classical novae limit cycle. 
In addition to self-consistently finding the range of
$T_{\rm eff}$ as a function of $\langle \dot M \rangle$ during the cycle, we also
self-consistently find the ignition masses. Comparing these
theoretical masses to the observed ejected masses will tell us whether
the WD mass in CV's is secularly increasing or decreasing.  We close
by comparing our results to the accumulated observations of quiescent
DN and making predictions for the colors of low $\langle \dot M \rangle$ CV's in
quiescence that are applicable to searches for faint CVs in the field
and galactic globular clusters.

\end{abstract}

\section{Introduction} 

 Dwarf Novae (DN) systems contain a white dwarf (WD) accreting matter
at time-averaged rates $\langle \dot M \rangle<10^{-9}M_\odot \ {\rm yr}^{-1}$ from a
low-mass ($<0.5M_\odot$ typically) stellar companion. At these
$\langle \dot M \rangle$'s, the accretion disk is subject to a thermal instability
which causes it to rapidly transfer matter onto the WD (at $\dot M \gg
\langle \dot M \rangle$) for a few days to a week once every month to year.
The orbital periods of these binaries are usually
less than 2 hours (below the period gap), but there are also DN above
the period gap, $>$ 3 hours (see Shafter 1992).  The $\dot M$ onto the
WD can be low enough between outbursts that the optical/UV emission is
dominated by the internal luminosity of the WD, not the accretion
disk. Recent HST/STIS spectroscopy has spectrally resolved the WD's
contribution to the quiescent light and found effective
temperatures $T_{\rm eff}\sim 10,000-40,000 \ {\rm K}$
(see our Figure \ref{fig:lars} and Sion 1999).

 The measured internal WD luminosity is larger than expected from an
isolated WD of similar age ($\approx$ Gyr), indicating that it has
been heated by accretion (Sion 1985).  Compressional heating
(i.e. internal gravitational energy release) appears to be the main
driver for this re-heating (Sion 1995). Sion's (1995) early estimate
for internal gravitational energy release within the WD (of mass $M$
and radius $R$) was $L\approx 0.15 GM\langle \dot M \rangle/R$. However, we show in \S
2 that most energy is released in the accreted outer envelope, giving 
$L\approx 3kT_c \langle \dot M \rangle/\mu m_p$, where $\mu\approx 0.6$ is
the mean molecular weight of the accreted material, $m_p$ is the
baryon mass, and $k$ is Boltzmann's constant. The theoretical
challenge that we address in \S 3 is how to calculate the WD core
temperature, $T_c$, as a function of $\langle \dot M \rangle$, and thus find $T_{\rm
eff}$. Because of the unstable nuclear burning and resulting classical
novae cycle, the envelope mass changes with time. This allows the core
to cool at low accumulated masses and be heated prior to unstable
ignition. We use nova ignition to determine the maximum mass of the
overlying freshly accreted shell, and find the steady-state (i.e. cooling
equals heating throughout the classical novae cycle) deep
interior temperature of the accreting WD, $T_c$, as a function of
$\langle \dot M \rangle$ and WD mass.

  In \S 4, we compare our theoretical work to STIS observations and
infer $\langle \dot M \rangle$ on the timescale of $10^6$ years, critical to
constraining CV evolutionary models.  Figure \ref{fig:lars} shows that
DN above the period gap are hotter than those below the gap, and have
$\langle \dot M \rangle$'s consistent with that expected from traditional CV evolution
(e.g. Howell et al. 2001), even those that involve some
``hibernation'' (Shara et al. 1986; Kolb et al. 2001). The result is
more surprising if the much weaker magnetic braking laws of Andronov
et. al. (2001) are correct. We also predict the minimum light ($M_V$)
of $\langle \dot M \rangle<10^{-10} M_\odot \ {\rm yr}^{-1}$ CVs in quiescence,
allowing for discovery of the predicted large population of CVs with
very low mass companions ($<0.1M_\odot$) that are near the period
minimum (Howell et al. 1997). Observations already show that the WD
fixes the quiescent colors of these CVs and our calculations are
useful for surveys in the field that were discussed at this meeting
(e.g. 2DF, SDSS, see Marsh et al. 2001 and Szkody et al. 2001
contributions here), as well as HST CV searches in globular clusters.

\section{The Basic Physics of Compressional Heating}
\label{sec:physics}

  It is important to make clear what is meant by internal
gravitational energy release or compressional heating {\it within the
white dwarf.} Most are familiar with the energy released ($GMm_p/R$) when
a baryon falls from a large distance to the stellar surface. This
energy is deposited at, or near, the photosphere and is rapidly
radiated away. This energy does not get taken into the star, as in
the upper atmosphere (where $T\ll T_c$) the time it takes the fluid to
move inward is always much longer (by a factor of at least $T_c/T$)
than the time it takes for heat to escape. Thus, once accretion has
shut off, or diminished (such as in DN quiescence), this energy
release is no longer relevant. What is relevant is energy release deep
in the WD due to compression by the freshly accreted material and
hydrostatic nuclear burning. That energy takes a long time to exit,
will still be visible when accretion has halted, and sets $T_{\rm
eff}$ in quiescence.

  Let's start with the simplest view of compressional heating. In the
non-degenerate outer atmosphere, the fluid is slowly falling down in
the WD gravitational field, $g=GM/R^2$. A fluid element falls a
distance of order the scale height, $h=kT/\mu m_p g$, in the time it
takes to replace it by accretion, giving $L\sim \langle \dot M \rangle g h\sim \langle \dot M \rangle
k T/\mu m_p$. This exhibits the scaling that appears in the formal
viewpoint, which is to consider the heat equation
\begin{equation}
\label{eq:heateq}
T\frac{ds}{dt}=T\frac{\partial s}{\partial t}+T\vec v \cdot \nabla s=
-\frac{\partial L}{\partial M_r}
+\epsilon_N,
\end{equation} 
where  $\epsilon_N$ is the nuclear burning rate, $s$ is the
entropy, and $\vec v=-\langle \dot M \rangle \hat r/4\pi r^2 \rho$ is the slow
downward advection speed from accretion. In pressure units, and
neglecting nuclear burning and the time-dependent piece, this becomes 
\begin{equation} 
L=-\langle \dot M \rangle \int_0^P T \frac{\partial s}{\partial P} dP. 
\end{equation}
The entropy decreases inward (i.e. the envelope and core are not
convective), so this is an outward going $L$.  The entropy profile is
fixed by the temperature gradient needed to carry the luminosity
outward and thus we simultaneously solve equation (2) (putting back in
$\epsilon_N$) with the heat transport equation, using opacities and
conductivities from Iglesias \& Rogers (1996) and Itoh et
al. (1983). For an analytic understanding, we integrate through the
non-degenerate envelope, where $s=k\ln(T^{3/2}/\rho)/\mu m_p$. Here,
$L$ is nearly constant, giving $T^{8.5}\propto P^2$. Integrating down
to the isothermal core yields $L\approx 3kT_c \langle \dot M \rangle/\mu m_p$.

 Now, let's turn to the degenerate carbon/oxygen core. For the
$\langle \dot M \rangle$'s and typical $M=0.6 M_\odot$ WD of interest here, all of the
entropy is in the liquid ions at $T_c\approx 10^7 \ {\rm K}$. The time
it takes to transport heat through the interior is $\sim 10^7 {\rm
yr}\ll M/\langle \dot M \rangle$, so the core is isothermal and any compression is far
from adiabatic.\footnote{This is in contrast to the rapid accretion
rates $\langle \dot M \rangle \gg 10^{-8} M_\odot \ {\rm yr}^{-1}$ considered for more
massive Type Ia progenitors, where the interior undergoes nearly
adiabatic compression (see Bravo et al. 1996).}  Due to uncertainty
from the classical novae cycle, we don't know whether the C/O core is
secularly increasing in mass, but if it were, almost all of the work
of compression goes into increasing the electron Fermi energy.  The
integrated heat release would only be $L\approx 15 kT_c\langle \dot M \rangle/\mu_i
m_p$ (Nomoto 1982) for a $0.6M_\odot$ C/O WD, where $\mu_i\approx 14$
is the ion mean molecular weight. Because of the mean molecular weight
contrast, this is about a factor of five smaller than that released in
the envelope. Thus, the entropy drop through the accreted layer is
larger than that across the core, making the accreted layer the main
source of compressional heating. 
 
\section{ Finding the Equilibrium Core Temperature}
\label{sec:Tcmethod}

  For our initial study, we dropped the time-dependent term in
equation (\ref{eq:heateq}), and presumed that the C/O WD mass was
constant throughout the classical nova cycle, thus only accounting for
the compressional heating and $\epsilon_N$ in the H/He
layer. This method improves on that of Iben et al. (1992) by allowing
the accreted envelope mass to change through the $10^5$ year classical
nova cycle. Early in the cycle, the mass of the accreted layer
is small, compressional heating is small, and the WD cools. Later in
the cycle, the accreted layer becomes thick enough that compressional
heating along with slow hydrogen burning releases a sufficient amount
of energy to heat the core.  As the WD has a large heat capacity,
reaching the equilibrium $T_c$ where the heat exchanged between the
envelope and core averages to zero over a single classical nova cycle
takes $\approx 10^8$ years. Since this time is usually shorter than
the time over which $\langle \dot M \rangle$ changes, we construct such equilibrium
accretors for a given $M$ and $\langle \dot M \rangle$.

\begin{figure}
\plotfiddle{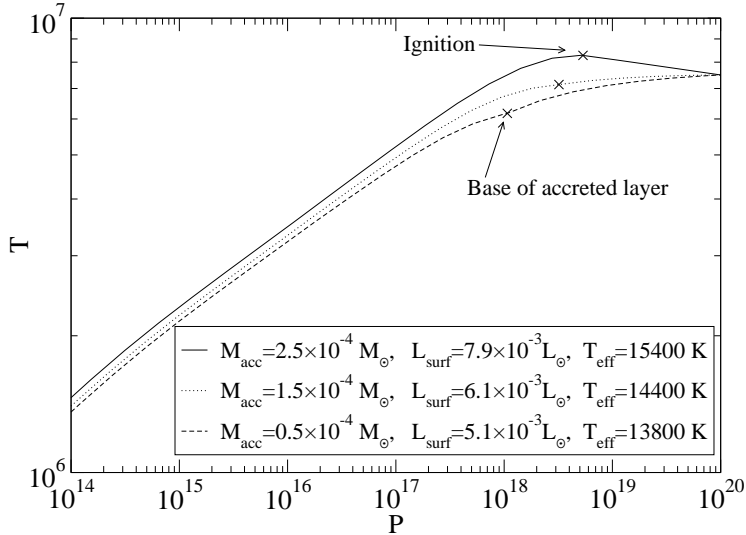}{2.6in}{0}{40}{40}{-160}{-20}
\caption{\label{fig:T-P} The Hydrogen/Helium envelope
and outer core in temperature and pressure for three different values
of accumulated mass, $M_{acc} = 0.5\times 10^{-4}M_\odot$, $1.5\times
10^{-4}M_\odot$ and $2.5\times 10^{-4}M_\odot$ for $M=0.6M_\odot$ and
$\langle \dot M \rangle=10^{-10}M_\odot$/yr.  The external surface luminosity of the
WD and the corresponding $T_{\rm eff}$ is also listed.  The part of
the star off to the right of the figure ($P>10^{20}$ dyne cm$^{-2}$)
is the isothermal inner core.}
\vspace{-20pt}
\end{figure}

 To do this, we first fix $T_c$ at the outer edge of the C/O core at a
pressure high enough so that the changing accumulated mass has little
direct effect.  With a radiative outer boundary condition, we then
integrate our structure equations with equation (\ref{eq:heateq}) to
find the envelope state for an $\langle \dot M \rangle$ and accreted layer mass.  See
Figure \ref{fig:T-P} for examples of the resulting $T$-$P$ relations.
We then evaluate the luminosity across the chosen location (the right
edge of the plot in Figure \ref{fig:T-P}) for a number of different
accreted layer masses up to the unstable ignition. The ignition mass
is found by comparing the $T$ and $\rho$ at the base of the accreted
(hydrogen rich) layer with the analytic ignition curves in Fujimoto
(1982). 

\begin{figure}
\plotfiddle{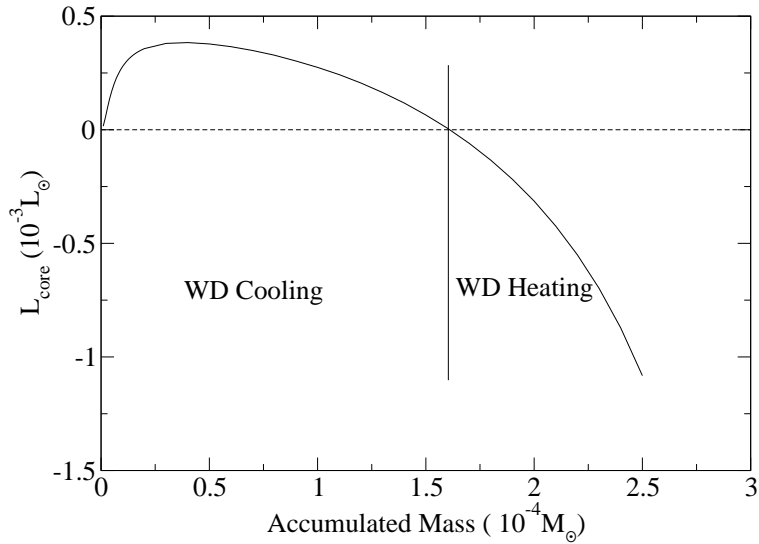}{2.6in}{0}{40}{40}{-160}{-20}
\caption{\label{fig:loop}
Luminosity at the outer edge of the core as
a function of accumulated mass up to the classical nova ignition for
the equilibrium model of Figure \ref{fig:T-P}. 
Positive is outgoing and epochs of core
cooling and heating are indicated.}
\end{figure}

\begin{figure}
\plotfiddle{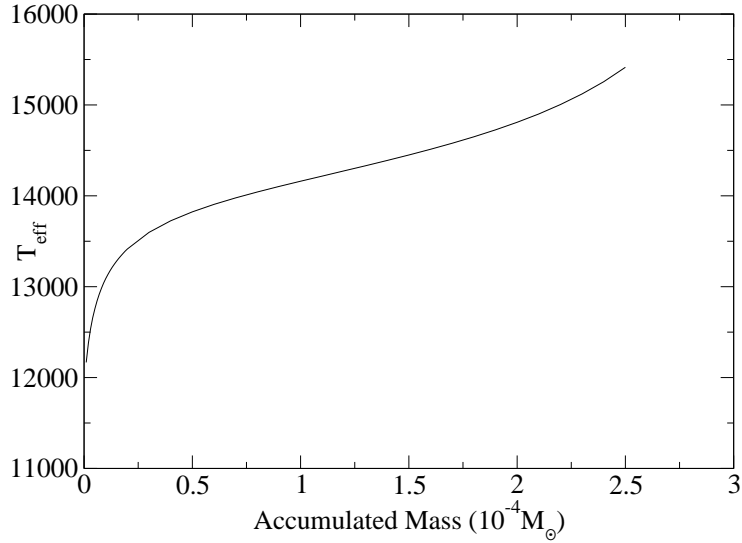}{2.6in}{0}{40}{40}{-160}{-20}
\caption{ \label{fig:Teff} Effective temperature
of the WD as a function of
accumulated mass for equilibrium model of Figure \ref{fig:T-P}.}
\end{figure}

We vary $T_c$ to find an equilibrium model, where the ``core
luminosity'' ($L_{\rm core}$) averages to zero over the classical nova
cycle as
shown in Figure \ref{fig:loop}. The quiescent $T_{\rm eff}$ which is
expected for the same cycle is shown in Figure \ref{fig:Teff}. At the
nova outburst we assume that the accreted shell is expelled, and that
due to the rapidity of this event, it does not appreciably heat the
WD. The resulting equilibrium core temperatures for
$\langle \dot M \rangle=10^{-10}M_\odot/$yr are $T_c=9\times 10^6 \ {\rm K}, 7.5\times
10^6 \ {\rm K}$ and $8.5\times 10^6 \ {\rm K}$ for $M=0.4, 0.6$ and
$1.0M_\odot$.  The $M=0.4M_\odot$ star is hotter than the
$M=0.6M_\odot$ star because it has a larger maximum accumulated mass
that leads to a longer period of core heating. For a $0.6M_\odot$ WD,
the core temperatures are $T_c/10^6{\rm K}=4,5.3,12.2$ and 18.0
for $\langle \dot M \rangle/M_\odot \ {\rm yr^{-1}}=10^{-11},3.2\times 10^{-11},
4.2\times 10^{-10}$ and $10^{-9}$. This also gives us
the $T_{\rm eff}$ range during the classical nova cycle, which we now
compare to observations.

\section{Comparison with Observations}

\begin{figure}
\plotfiddle{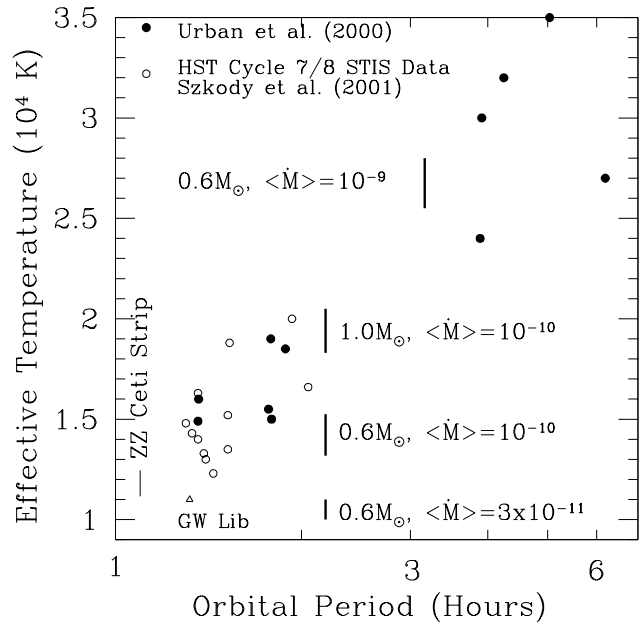}{3.0in}{0}{45}{45}{-135}{-80}
\caption{\label{fig:lars} 
The quiescent  WD $T_{\rm eff}$'s in DN as a function of binary orbital period.
The open circles are Cycle 7/8 STIS observations from Szkody et. al. (priv.
commun.), while the solid circles are from the compilation of Urban et al.
(2000). The single point for GW Lib is from Szkody et al. (2000), which is cold
enough ($T_{\rm eff}\approx 11,000 \ {\rm K}$; Van Zyl 1998) to be a ZZ Ceti
pulsator in quiescence. The thick vertical lines are our current predictions
for the $T_{\rm eff}$ range for the stated masses and $\langle \dot M \rangle$'s. The thin
solid line is the ZZ Ceti instability strip (Bergeron et al. 1995).}
\end{figure}

  In contrast to Sion's (1995) simulations which only directly
addressed the immediate cooling after outburst, our work has focused
on the release of heat from the deep interior.  The long-timescale
nature of heating in these deep layers allows a comparison between our
results and the observed $T_{\rm eff}$'s.  For CVs in the field, the
large set of STIS observations by Szkody et al.\ (2001) and previous
observations (Urban et al. 2000) provide
detailed spectra of quiescent WDs in DN. Figure \ref{fig:lars} shows
the current $T_{\rm eff}$ measurements and compares them to our
results. The measurements are made during deep quiescence when the
accretion luminosity is negligible and are intended to be long enough
after the outbursts that other emission mechanisms (e.g. Pringle's
(1988) suggestion of radiative illumination of the WD) have faded. In
that case, we are observing heat directly from the deep WD interior.
The number of such systems will increase due to upcoming all-sky
surveys (such as SDSS and 2DF, see Marsh et al. 2001), and will push
to lower $\langle \dot M \rangle$ systems with long quiescent intervals.

  We can already give some general insight from the initial results
shown. Our work indicates that below the period gap, $\langle \dot M \rangle\approx
10^{-10}M_\odot\ \rm yr^{-1}$ and the WD masses are in the range
$0.6$-$1.0M_\odot$. This agrees with the expectation from Kolb \&
Baraffe (1999), who find $\langle \dot M \rangle\approx 5\times 10^{-11} M_\odot \
{\rm yr}^{-1}$ at an orbital period of 2 hours presuming angular
momentum losses from gravitational waves alone. Above the period gap, the
$T_{\rm eff}$ is higher, and we estimate $\langle \dot M \rangle\approx
10^{-9}M_\odot\ \rm yr^{-1}$.

We predict that a $0.6M_\odot$ WD above the gap has a core
temperature of $1.8\times 10^7$ K and, if in equilibrium below the
gap, $T_c=7.5\times 10^6$K.  An interesting question is whether the WD
will have time to cool this much as it traverses the gap. If not, then
the WD's below the period gap will be hotter than our calculation
implies. We estimate this cooling time from the current  WD cooling
law (e.g. Chabrier et al. 2000) along with the heat capacity of the core,
\begin{equation} 
L_{\rm cool} \approx 10^{-2}L_\odot \left(\frac{T_c}{1.8\times
10^7 {\rm K}}\right)^{2.5}=
-\frac{d}{dt}\left[\frac{3k_BT}{\mu_im_p}M\right], 
\end{equation} 
which gives $\Delta t\approx 0.5$ Gyr. Since this is comparable to the
estimated time spent in the gap (Howell et al. 2001), our equilibrium
assumption below the gap is likely safe. However, note that about 0.2
Gyrs after accretion halts, the WD will enter the ZZ-Ceti instability
strip!

These results also have bearing on the search for faint CVs in
globular clusters. The expected population might well contain many low
$\langle \dot M \rangle$ systems that spend much of their time in quiescence. These
CVs are commonly searched for via the presence of hydrogen emission
lines or X-ray emission (as recent Chandra observations have found;
Grindlay et al. 2001a, Grindlay et al. 2001b), and this method is
fruitful. However, for the very-low $\langle \dot M \rangle$ systems, the disks can be
very dim and the quiescent X-ray emission too faint even for
Chandra. We now show that these systems (as well as CVs crossing the
period gap or those ``hibernating'' post-novae, Shara et al. 1986) can
be identified by their position in a color-magnitude diagram (CMD). By
using our theory of the thermal state of the WD, it is possible
to predict the broadband colors of CV systems without dependence on
the disk luminosity.

\begin{figure}
\plotfiddle{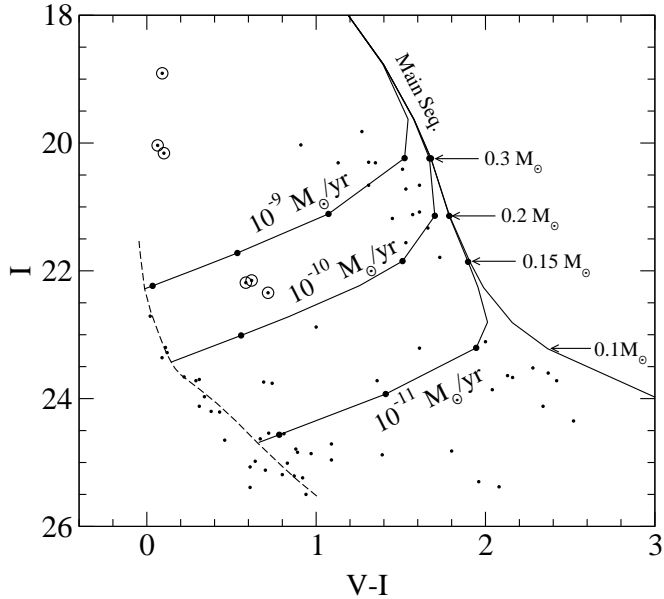}{3.0in}{0}{45}{45}{-140}{-20}
\caption{\label{fig:gc} Color-Magnitude Diagram of NGC 6397.  The dots
are members of the cluster that are below the MS (HST observations by
King et al. 1998) and the $\odot$'s are the ``Non-Flickerers'' from
Taylor et al.  (2001). The MS line is from Baraffe et al. (1997) for
[M/H]$=-2.0$ and an age of 10 Gyr (actual cluster [M/H]$=-1.95$). The
dashed line is from Bergeron et al. (1995) for DA WDs with $\log
g=8$. The lines connecting the MS to the WD sequence are our current
calculations of the WD+MS binary at the specified $\langle \dot M \rangle$. The
highest $T_{\rm eff}$ during the classical nova cycle has been used
for the WD in each case (see Figure \ref{fig:Teff}). No disk has been
included, which is a safe guess for the low $\langle \dot M \rangle$ systems. All
curves have been put at the distance and reddening of the
cluster, $(m-M)_I=12.05$ and $E(V-I)=0.288$.}
\end{figure}

 An excellent example is NGC 6397 (King et al. 1998; Taylor et
al. 2001). Figure \ref{fig:gc} shows a CMD of NGC 6397 with our
initial results that provide a relationship between $T_{\rm eff}$ and 
$\langle \dot M \rangle$. The lines were produced by superposing a WD with
the maximum $T_{\rm eff}$ for the indicated $\langle \dot M \rangle$ with a MS
star. Except for near the WD cooling line (dashed curve), where the WD
becomes completely dominant, the $I$ magnitude is set by the MS
companion.  The large dots along the $10^{-9}M_\odot \ {\rm yr}^{-1} $
and $10^{-10}M_\odot \ {\rm yr}^{-1}$ lines indicate where the MS
companion is 0.3, 0.2, 0.15 and 0.1 $M_\odot$, and two additional
points at 0.09 and 0.085 $M_\odot$ are indicated on the
$10^{-11}M_\odot \ {\rm yr}^{-1}$ line. This immediately provides a
number of candidate systems (namely, those data residing in this part
of the CMD).

 The circled points are the ``non-flickerers'' (Cool et. al. 1998)
recently reported by Taylor et al. (2001). The three at $I\approx
22.25$ are very strong H$\alpha$ absorbers (consistent with a DA WD)
and were not detected by Chandra (Grindlay et al. 2001b). These
authors had discussed these systems as possible helium WDs with
millisecond pulsar companions, though, given our work, we would claim
that these are hot WDs with $\approx 0.15 M_\odot$ MS companions. 
In addition, the population of data points in this diagram with respect
to our theoretical curves will eventually constrain CV evolutionary
scenarios.  If we assume many of the data points are CVs, we already
see that most systems with high $\langle \dot M \rangle$ have $0.15-0.3 M_\odot$
companions. The large number of data points below the
$\langle \dot M \rangle=10^{-11}M_\odot \ {\rm yr}^{-1}$ line could well be the
long-sought post-turnaround systems with $\langle \dot M \rangle=10^{-12}M_\odot \
{\rm yr}^{-1}$ and companion masses $< 0.09 M_\odot$ (Howell et
al. 1997).

\section{Conclusion and Future Directions}

We have evaluated the action of compressional heating of accreting WD
interiors. Most of the compressional energy release takes place in the
accreted envelope, and is thermally communicated to the core. The
maximum envelope mass is set by the unstable nuclear burning that
causes a classical nova runaway and most likely expels the accreted
mass. We have constructed equilibrium accretors which have constant
core temperatures such that the heat lost from the core when the
envelope is thin (i.e. right after the classical nova) is balanced by
that regained when the envelope is thick. This equilibrium determines
the $T_{\rm eff}$ of the WD throughout the classical nova cycle.  Our
models agree with the observations of Dwarf Novae in deep quiescence
and imply $\langle \dot M \rangle\approx10^{-10}M_\odot$ yr$^{-1}$ just below the
period gap and $\langle \dot M \rangle\approx10^{-9}M_\odot$ yr$^{-1}$ just above the
period gap for WD masses in the range $0.6$--$1.0M_\odot$.

 Though our initial efforts have met with apparent success, there is
still much to be done. First we need to investigate the relevant
parameter ranges (e.g. $M$ and $\langle \dot M \rangle$) within our initial scheme.
The most critical parameter to vary is the metallicity of the accreted
material, lowering to values relevant for globular cluster science. We
must also survey our initial assumptions:

\begin{itemize}
\item WD excavation or accretion. The assumption that there was no net
mass loss or gained by the WD through the classical nova cycle must be
relaxed. This will allow for cooling of the WD due to adiabatic
expansion after mass loss, or heating if it is increasing in mass. The
large C/O fractions seen in some novae ejecta (Gehrz et al. 1998)
might indicate that the WD is decreasing in mass at $\langle \dot M \rangle$.

\item A self consistent accounting for the ignition masses, including
varying the metallicity. Our initial work used Fujimoto's (1982)
results for simplicity, but these are limited at low 
$\langle \dot M \rangle$'s.
A careful comparison to more modern nova calculations (e.g. Prialnik \& Kovetz
1995) must be carried out.

\item Thermal evolution of the WD. We also need to concern ourselves with
the secular change of $\langle \dot M \rangle$ due to the decreasing companion mass or
changing angular momentum losses (such as a drop in magnetic
braking at the period gap). When this occurs on a timescale comparable
to the WD thermal time, it is possible that the WD will not reach the
steady-state solution we have assumed. This memory of previous  higher
$\langle \dot M \rangle$ epochs could well allow the WD to be hotter than the
equilibrium accretor.
\end{itemize}

   While these steps are unlikely to change our understanding of the
compressional heating mechanism, each is essential for applying our
work to the observations. As assumptions are relaxed and investigated,
much more will be learned about both the state of CV systems and their
evolution.

\acknowledgements 

 We thank Paula Szkody for sharing the most recent HST observations.
 This research was supported by NASA via grant NAG 5-8658 and by the
 NSF under Grants PHY99-07949 and AY97-31632. L. B. is a Cottrell
 Scholar of the Research Corporation.

\end{document}